\documentclass[%
 reprint,
superscriptaddress,
 amsmath,amssymb,
 aps,
 pre,
]{revtex4-1}

\usepackage{amssymb,amsmath,bm}
\usepackage{enumerate}
\usepackage[utf8]{inputenc}
\usepackage{cancel}
\usepackage{ulem}
\usepackage{color}
\usepackage[colorlinks,linkcolor=blue,urlcolor=blue,citecolor=blue]{hyperref}
\usepackage{xr}
\usepackage[toc]{appendix}
\usepackage{graphicx,epstopdf}
\usepackage{tikz}

\def\be{\begin{equation}}
\def\ee{\end{equation}}
\def\bea{\begin{eqnarray}}
\def\eea{\end{eqnarray}}

\newcommand{\eref}[1]{Eq.~(\ref{#1})}%
\newcommand{\fref}[1]{Fig.~\ref{#1}} %
\begin{document}
\title{Universal criterion for selective outcomes under stochastic resetting}
\author{Suvam Pal}
\email{suvamjoy256@gmail.com}
\affiliation{Physics and Applied Mathematics Unit, Indian Statistical Institute, 203 B.T. Road, Kolkata 700108, India}
\author{Leonardo Dagdug}
 \email{dll@xanum.uam.mx}
\affiliation{Physics Department, Universidad Aut\'onoma Metropolitana-Iztapalapa, San Rafael Atlixco 186, Ciudad de M\'exico, 09340, M\'exico.}
\author{Dibakar Ghosh}
\email{dibakar@isical.ac.in}
\affiliation{Physics and Applied Mathematics Unit, Indian Statistical Institute, 203 B.T. Road, Kolkata 700108, India}
 \author{Denis Boyer}
 \email{boyer@fisica.unam.mx}
 \affiliation{Instituto de F\'isica, Universidad Nacional Aut\'onoma de M\'exico, Ciudad de M\'exico C.P. 04510 M\'exico.}
 \author{Arnab Pal}
 \thanks{Corresponding author}
\email{arnabpal@imsc.res.in}
\affiliation{The Institute of Mathematical Sciences, CIT Campus, Taramani, Chennai 600113, India \& 
Homi Bhabha National Institute, Training School Complex, Anushakti Nagar, Mumbai 400094,
India}

\begin{abstract}
    Resetting plays a pivotal role in optimizing the completion time of  complex \textcolor{black}{first-passage} processes with single or multiple outcomes/exit possibilities. While it is well established that the coefficient of variation -- a statistical dispersion defined as a ratio of the fluctuations over the mean of the \textcolor{black}{first-passage} time -- must be larger than unity for resetting to be beneficial for any outcome averaged over all the possibilities, the same can not be said while conditioned on a particular outcome. The purpose of this letter is to derive a universal condition which reveals that two statistical \textcolor{black}{metrics} -- the mean and coefficient of variation of the conditional times -- come together to determine when resetting can expedite the completion of a selective outcome,  and furthermore can govern the biasing between preferential and non-preferential outcomes. The universality of this result is demonstrated for a one dimensional diffusion process subjected to resetting with two absorbing boundaries. 
\end{abstract}

\maketitle

\section{Introduction}

Stochastic resetting has gained a lot of interest over the last decade due to its cross-disciplinary applications in physics \cite{evans_diffusion_2011,evans2020stochastic,evans2013optimal,pal2015diffusion,kumar2023universal,pal2017first,reuveni_optimal_2016,chechkin2018random,belan2018restart,eule2016non,barman2025optimizing}, chemistry \cite{reuveni2014role,biswas2023rate}, biology \cite{roldan2016stochastic,budnar2019anillin}, operation research \cite{bonomo2021mitigating,roy2024queues}, economics \cite{jolakoski2023first,stojkoski2022income} and ecology \cite{boyer2014random,paramanick2024uncovering}.
\textcolor{black}{Resetting is a simple mechanism where a stochastic dynamical system is stopped  at random times and is reset instantaneously to its initial configurations.}
A hallmark feature of resetting is its ability to expedite the completion of search processes. \textcolor{black}{In complex first-passage processes, the underlying dynamics can lead to significant fluctuations, often resulting in inefficient search behavior. To mitigate this, a variety of random search strategies have been developed—many inspired by natural phenomena observed in biology, chemistry, epidemiology, ecology, and finance—to improve navigational efficiency in engineered systems \cite{metzler2014first, holcman2014narrow, benichou2011intermittent,benichou2008optimizing,bray2013persistence, redner2001guide, grebenkov2024target}. Among these, stochastic resetting has emerged as a particularly effective strategy. By eliminating errant trajectories and enabling re-exploration of the phase space, resetting can significantly optimize the mean search time for locating targets \cite{evans2020stochastic,pal2024random,gupta2022stochastic}}. Naturally, the question arises on the universal performance of resetting dynamics and the quantification of physical conditions which can underpin the robustness beneath this phenomena. Recently, this question was addressed and it was shown that \textcolor{black}{the mean first-passage time of any process can be decreased by stochastic resetting (at constant rate)} if a $CV$-criterion is fulfilled \cite{reuveni_optimal_2016,pal2017first}, namely 
\begin{equation}\label{cv}
    CV\left(\mathbf{\Sigma}\right)>1,
\end{equation} 
where $CV$, the coefficient of variation, is defined as a ratio between the standard deviation and the mean of the search time for the system of interest in the absence of resetting, and $\mathbf{\Sigma}$ is a set of relevant system parameters such as the initial configurations and other specifics that impact the search time. The universality of this condition is quite striking since it neither depends on the exact nature of the dynamics nor on the dimensionality of the system. On a more physical ground, this condition essentially indicates that for underlying processes with broad \textcolor{black}{first-passage} time distributions, the effect of resetting is more pronounced. The ubiquity of this criterion hitherto has been verified in various set-ups. We refer to \cite{pal2024random,pal2022inspection} for an illustrative discussion on this criterion.

\begin{figure}
\centering
\includegraphics[width=\columnwidth]{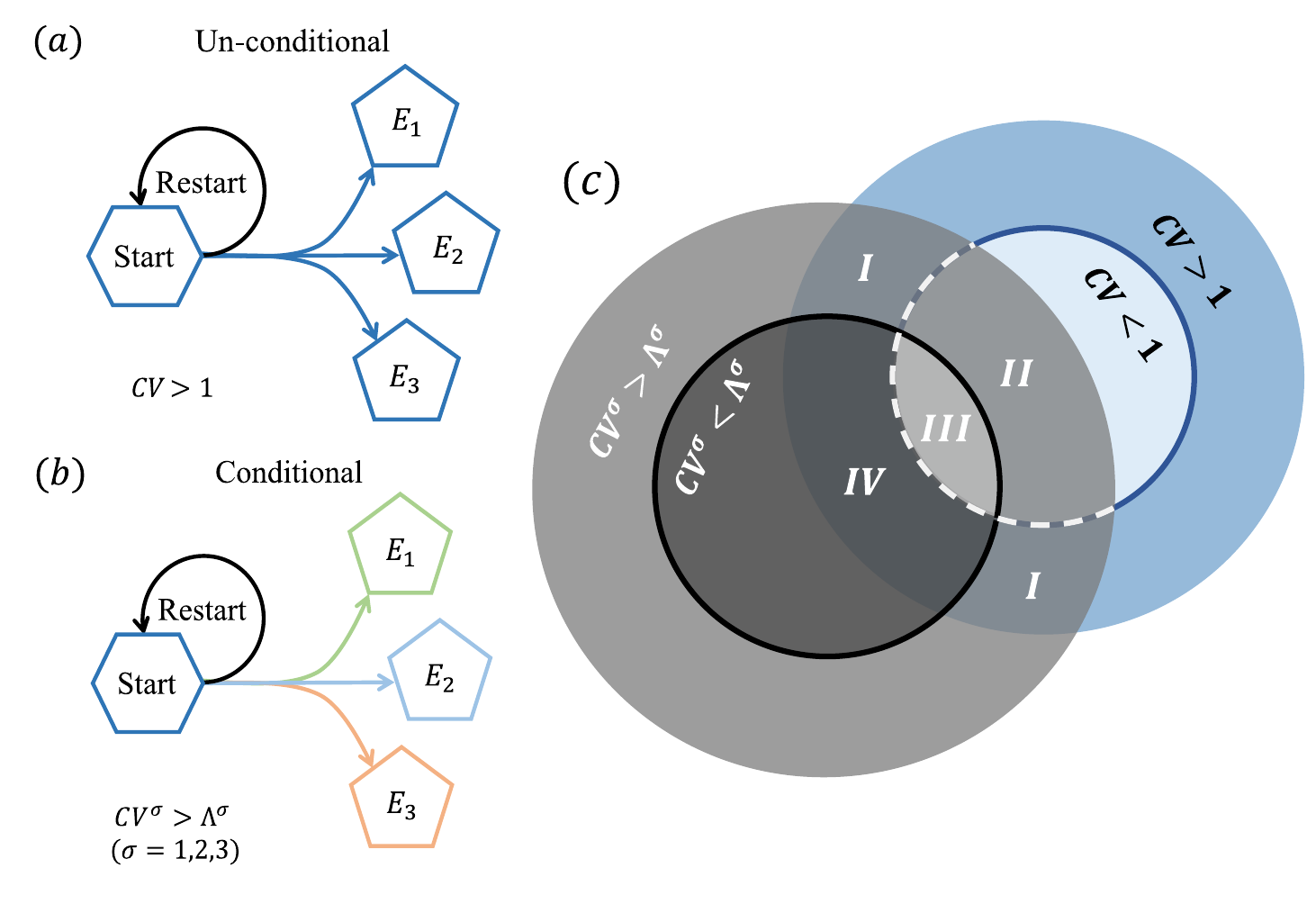}
    \caption{Resetting mediated universal criterion for selective and non-selective outcomes for a \textcolor{black}{first-passage} process with multiple exits or outcomes labeled as $E_1, E_2$, $E_3$ etc. Panel (a) illustrates the unconditional completion of the process (regardless of the outcome) averaged over all the possibilities. In contrast, panel (b) shows the conditional completion associated with a specific outcome (indicated by their respective color). Panel (c) uses a Venn diagram to illustrate the universal criteria and their domain of validity marking, in particular, the trade-off boundary as in \eref{lambda-c} and various regions of competing outcomes (as explained in Sections \ref{formalism:universal_conditional} and \ref{formalism:trade-off}).}
    \label{fig1}
\end{figure}

To motivate this work, let us now consider a complex \textcolor{black}{first-passage} time process with multiple targets or outcomes, indexed by $\{\sigma\}$. \textcolor{black}{Absorption processes in the presence of several targets or exit points in the sample space have been well studied for simple diffusion or intermittent processes, and have also been discussed in the context of resetting} \cite{redner2001guide, coppey2004kinetics,pal2019first,pal2019landau,bressloff2020directed,bressloff2020modeling,bressloff2020target,chechkin2018random,belan2018restart,ahmad2019first,bonomo2021first,jain2023fick,pal2024channel}. 
\textcolor{black}{A representative example arises in protein–DNA interactions, where proteins search for specific target sites along a DNA strand \cite{coppey2004kinetics,benichou2011intermittent,mirny2009protein}. In such systems, two target sites may compete for reaction with the same enzyme molecule, which undergoes 1D diffusion along the DNA interspersed with intermittent 3D relocations to random positions on the strand \cite{coppey2004kinetics}. This intermittent search mechanism closely resembles stochastic resetting to random locations with overhead times \cite{benichou2011intermittent,mirny2009protein}. Moreover, the presence of multiple target sites naturally introduces the need to consider conditional and selective reaction probabilities. This motivates us to consider searches conditioned on one or a set of particular targets (selective outcomes) instead of focusing on the unconditional \textcolor{black}{first-passage} time of a resetting process, regardless of the specific choice of the target (hence, non-selective outcome), as illustrated in \fref{fig1}.} 
Quite interestingly, 
while the above $CV$-criterion puts a strong constraint on the non-selective outcome, it cannot predict the effect of restart towards a selective outcome. The central purpose of this letter is to reveal a universal criterion for the faster completion (\textcolor{black}{in the level of mean search time}) of a conditional outcome $\sigma$ via resetting,
\begin{equation}\label{main}
CV^\sigma\left(\mathbf{\Sigma}\right)>\Lambda^\sigma\left(\mathbf{\Sigma}\right),
\end{equation}
that involves conditional and unconditional means, as well as the $CV$'s of the \textcolor{black}{first-passage} time for selective outcomes (reflected onto $\Lambda^{\sigma}$ to be defined later). 
We also deduce how resetting can bias the system towards a desired outcome out of many in comparison to a non-selective outcome. These universal frontiers are then illustrated rigorously for a diffusion process.

\section{General formalism}

Let us consider a stochastic search process confined in a  \textcolor{black}{$d$-dimensional} domain $\mathbf{\Omega}$ with multiple exit possibilities, labeled with the set-index $\{ \sigma \}$ \textcolor{black}{(see \cite{coppey2004kinetics} for such an instance where an enzyme protein explores various stochastic pathways to detect competitive target sites on a long DNA)}. The search process is intermittently stopped and reset to its initial set-up at rate $r$. We focus on two key observables: (i) $T_r$ -- the unconditional random time of completion regardless of the specific target attained, (ii) $T_r^\sigma$ -- the random time of completion conditioned on a specific target $\sigma$. We will demonstrate below that while \eref{cv} governs the optimization of $\langle T_r \rangle$, that of $\langle T_r^\sigma \rangle$ is determined by \eref{main}. 

\subsection{Currents and conditional escapes}
\label{formalism:currents}
We start by defining a probability current $J^{\sigma}_r\left(\mathbf{\Sigma},t\right)$ assigned with each outcome or target $\sigma$ up-to a time $t$. Since the system is reset at a rate $r$, we can write a renewal equation for the individual probability current,
\begin{align}\label{renewal-J}
    J^{\sigma}_r\left(\mathbf{\Sigma},t\right)&=e^{-rt}J^{\sigma}_0\left(\mathbf{\Sigma},t\right)\nonumber\\
    & + r\int_{0}^{t}d\tau~e^{-r\tau}Q_r\left(\mathbf{\Sigma},t-\tau\right)J^{\sigma}_0\left(\mathbf{\Sigma},\tau\right),
\end{align}
where $J^{\sigma}_0\left(\mathbf{\Sigma},t\right)$ is the probability current through $\sigma$ for the reset-free underlying process, and $Q_r\left(\mathbf{\Sigma},t\right)$ is the survival probability that the process has not been able to find any targets or exit points up-to time $t$. The rationale behind the renewal equation stems from two possible events, where (i) the process ended with no resetting events (the first term on the rhs), and (ii) the process ended experiencing multiple resetting events (the second term on the rhs). Similarly, the renewal relation for the survival probability is given by
$
Q_r\left(\mathbf{\Sigma},t\right)=e^{-rt} Q_0\left(\mathbf{\Sigma},t\right)+r\int_0^t d\tau e^{-r\tau} Q_0\left(\mathbf{\Sigma},\tau\right)Q_r\left(\mathbf{\Sigma},t-\tau\right)
$
\cite{evans2020stochastic,pal2019first}, where $Q_0\left(\mathbf{\Sigma},t\right)$ is the reset-free survival probability. Note that $Q_0\left(\mathbf{\Sigma},t\right)=\int_t^\infty~dt~f_{T_0}(t)$, where $f_{T_0}(t)$ is the unconditional \textcolor{black}{first-passage} time density for the underlying process. Taking Laplace transform of these two equations, we arrive at
\begin{align}\label{renewal-lap-J}
    \widetilde{J}_r^{\sigma}\left(\mathbf{\Sigma},s\right)=\dfrac{\widetilde{J}_0^{\sigma}\left(\mathbf{\Sigma},r+s\right)}{1-r\widetilde{Q}_0\left(\mathbf{\Sigma},r+s\right)},
\end{align}
where we have denoted the Laplace transform of a function $f(t)$ as 
$\widetilde{f}(s)=\int_0^{\infty}dt\ e^{-st} f(t)$.
%
It is evident that \eref{renewal-lap-J} allows us to connect the currents to the underlying process, thus reducing the complexity of the problem.

The conditional \textcolor{black}{first-passage} time density is given by \cite{redner2001guide,coppey2004kinetics}
\begin{align}
\label{FPT_r}
    f_{T_r^\sigma}(\mathbf{\Sigma},t)=\frac{J^{\sigma}_r\left(\mathbf{\Sigma},t\right)}{\epsilon_r^{\sigma}\left(\mathbf{\Sigma}\right)},
\end{align}
where $\epsilon_r^{\sigma}\left(\mathbf{\Sigma}\right)$ is the exit or splitting probability that the searcher arrives at a specific outcome $\sigma$ (e.g., exits via a specific target) before arriving elsewhere and is simply the integrated probability flux through that target \cite{redner2001guide,coppey2004kinetics}
\begin{equation}\label{epsilon}
\epsilon_r^{\sigma}\left(\mathbf{\Sigma}\right)=\int_{0}^{\infty}J^{\sigma}_r\left(\mathbf{\Sigma},t\right)~dt.
\end{equation}
The relation (\ref{FPT_r}) allows us to compute the moments of the conditional exit times through specific boundaries. For instance, the conditional mean \textcolor{black}{first-passage} time through a specific exit $\sigma$ is given by \cite{redner2001guide,coppey2004kinetics}
\begin{equation}\label{def-conditional-mfpt}
    \langle T^{\sigma}_r(\mathbf{\Sigma})\rangle=\int_{0}^{\infty} dt ~t  f_{T_r^\sigma}(\mathbf{\Sigma},t)=\dfrac{\int_{0}^{\infty} dt~t J^{\sigma}_r(\mathbf{\Sigma},t)}{\epsilon_r^{\sigma}\left(\mathbf{\Sigma}\right)},
\end{equation}
and the unconditional mean \textcolor{black}{first-passage} time is obtained by averaging over all the escape routes
\begin{equation}\label{conditional-mfpt-2}
    \langle T_r\left(\mathbf{\Sigma}\right)\rangle=\sum_{\sigma}\epsilon^\sigma_r\left(\mathbf{\Sigma}\right)\langle T^{\sigma}_r\left(\mathbf{\Sigma}\right)\rangle,
\end{equation}
while the unconditional \textcolor{black}{first-passage} time density can be written as $f_{T_r}(\mathbf{\Sigma},t)=-\partial Q_r(\mathbf{\Sigma},t)/\partial t$. The structure constituting Eqs. (\ref{renewal-lap-J})-(\ref{conditional-mfpt-2}) is generic and does hold for resetting and underlying processes with multiple outcomes or exit points. 

Utilizing Eq. (\ref{epsilon}), or $\epsilon_r^{\sigma}\left(\mathbf{\Sigma}\right)=\widetilde{J}^{\sigma}_r\left(\mathbf{\Sigma},s=0\right)$, and Eq. (\ref{renewal-lap-J}), we can represent the exit probabilities in terms of the underlying process 
\begin{equation}\label{epsilon-2}
    \epsilon_r^{\sigma}\left(\mathbf{\Sigma}\right)=\dfrac{\widetilde{J}_0^{\sigma}\left(\mathbf{\Sigma},r\right)}{1-r\widetilde{Q}_0\left(\mathbf{\Sigma},r\right)}=\dfrac{\epsilon_0^{\sigma}\left(\mathbf{\Sigma}\right) \widetilde{T}_0^\sigma\left(\mathbf{\Sigma},r\right)}{\widetilde{T}_0\left(\mathbf{\Sigma},r\right)}.
\end{equation}
In the second equality of Eq. (\ref{epsilon-2}), $\epsilon_0^{\sigma}\left(\mathbf{\Sigma}\right)$ is the splitting probability for the underlying process. Furthermore, $\widetilde{T}_0^{\sigma} \left(\mathbf{\Sigma},s\right)=\int_0^\infty dt~e^{-st}f_{T_0^\sigma}(\mathbf{\Sigma},t)=\langle e^{-sT_0^\sigma} \rangle$ and $\widetilde{T}_0 \left(\mathbf{\Sigma},s\right)=\int_0^\infty dt~e^{-st}f_{T_0}(\mathbf{\Sigma},t)=\langle e^{-sT_0} \rangle$ respectively are the conditional and unconditional \textcolor{black}{first-passage} time densities in the Laplace space for the underlying process. In Eq. (\ref{epsilon-2}), we have used the general relations $\widetilde{J}_0^{\sigma}\left(\mathbf{\Sigma},s\right)=\epsilon_0^{\sigma}\left(\mathbf{\Sigma}\right) \widetilde{T}_0^\sigma\left(\mathbf{\Sigma},s\right)$ [analogous to Eq. (\ref{FPT_r})] and $\widetilde{T}_0\left(\mathbf{\Sigma},s\right)=1-s\widetilde{Q}_0\left(\mathbf{\Sigma},s\right)$. With Eq. (\ref{epsilon-2}), we can now rewrite Eq. \eqref{def-conditional-mfpt} to obtain an expression for the conditional mean exit time [see Appendix \ref{appendixA} for details]
\begin{equation}\label{conditional-mfpt}
    \langle T^{\sigma}_r\left(\mathbf{\Sigma}\right)\rangle = \langle T_r\left(\mathbf{\Sigma}\right)\rangle+\frac{\partial}{\partial r} \ln \frac{\widetilde{T}_0\left(\mathbf{\Sigma},r\right)}{\widetilde{T}_0^\sigma\left(\mathbf{\Sigma},r\right)},
\end{equation}
which is represented in terms of the unconditional and conditional time statistics of the underlying process (see also \cite{belan2018restart,pal2019first}). This is an important relation that will be required to derive the universal criterion for resetting mediated conditional exits. Before doing so, we recall that the $CV$-criterion can be derived from the relation \cite{reuveni_optimal_2016,pal2017first}
\begin{eqnarray}\label{mfpt}
    \langle T_r\left(\mathbf{\Sigma}\right)\rangle=\widetilde{Q}_r\left(\mathbf{\Sigma},s=0\right)=\frac{1-\widetilde{T}_0\left(\mathbf{\Sigma},r\right)}{r\widetilde{T}_0\left(\mathbf{\Sigma},r\right)}.
\end{eqnarray}
Expanding Eq. (\ref{mfpt}) at first order near $\delta r=0$ and demanding $\langle T_{ \delta r}\left(\mathbf{\Sigma}\right)\rangle < \langle T_0\left(\mathbf{\Sigma}\right)\rangle$ results in 
\begin{equation}
CV\left(\mathbf{\Sigma}\right) \equiv \frac{\sqrt{\langle T_0(\mathbf{\Sigma})^2\rangle-\langle T_0(\mathbf{\Sigma})\rangle^2}}{\langle T_0(\mathbf{\Sigma})\rangle}>1,
\end{equation}
which turns out to be a sufficient condition for resetting to be beneficial for any unconditional exit.

\subsection{The universal criterion for conditional exits}
\label{formalism:universal_conditional}
To understand the effects of resetting on the conditional exit times, we perform a linear expansion of $\langle T_{\delta r}^{\sigma}\left(\mathbf{\Sigma}\right)\rangle$ in Eq. (\ref{conditional-mfpt}) near $\delta r=0$, as for the unconditional time. Following  Appendix \ref{appendixB}, one gets
\begin{align}\label{conditional-cv}
    \langle T_{\delta r}^{\sigma}\left(\mathbf{\Sigma}\right)\rangle&=\langle T^{\sigma}_0\left(\mathbf{\Sigma}\right)\rangle\nonumber\\
    &+\delta r~\langle T^{\sigma}_0\left(\mathbf{\Sigma}\right)\rangle^2\left[\Lambda^{\sigma}\left(\mathbf{\Sigma}\right)^2-CV^{\sigma}\left(\mathbf{\Sigma}\right)^2\right],
\end{align}
where 
$CV^{\sigma}\left(\mathbf{\Sigma}\right)=\sqrt{\langle T_0^{\sigma}(\mathbf{\Sigma})^2\rangle-\langle T_0^{\sigma}(\mathbf{\Sigma})\rangle^2}/\langle T_0^{\sigma}(\mathbf{\Sigma})\rangle
$ and $\Lambda^{\sigma}\left(\mathbf{\Sigma}\right)$ is a threshold given by
\begin{align} \Lambda^{\sigma}\left(\mathbf{\Sigma}\right)=\sqrt{\dfrac{\langle T_0\left(\mathbf{\Sigma}\right)\rangle^2}{2\langle T^{\sigma}_0\left(\mathbf{\Sigma}\right)\rangle^2}\left[1+CV\left(\mathbf{\Sigma}\right)^2\right]}. \label{Lambda_def}  
\end{align}
The above dimensionless quantity depends on both unconditional and conditional properties of the underlying process. Asking  $\langle T_{\delta r}^{\sigma}\left(\mathbf{\Sigma}\right)\rangle < \langle T_{0}^{\sigma}\left(\mathbf{\Sigma}\right)\rangle$ leads to the sufficient criterion \eqref{main} for resetting mediated conditional exits. This relation is universal and not specific to a system or a choice of the initial condition, or dimensionality, as long as the memory renewal holds after each resetting. 

 A careful observation of Eq. \eqref{main} and Eq. \eqref{Lambda_def} further indicates that for conditional escape to be beneficial, either of the following scenarios can be responsible: (i) the fluctuations of the conditional time $CV^{\sigma}$ be relatively larger than that $CV$ of the unconditional time, or (ii)  at fixed $CV^{\sigma}$ and $CV$, the mean conditional time be larger than the unconditional one, resulting in a smaller magnitude for the threshold $\Lambda^\sigma$. To understand this, imagine a scenario when a process starts its dynamics and can avail $\{ \sigma \}$-exit possibilities by choosing a large number of escape pathways. Biasing between the pathways can render large fluctuations in the conditional \textcolor{black}{first-passage} times leading to slow exits. Eq. \eqref{main} quantifies these aspects and in essence, suggests that resetting can be beneficial in such scenarios by intermittently cutting short the prolonged trajectories.

\subsection{Trade-off between resetting mediated unconditional and conditional exits}
\label{formalism:trade-off}
Various optimization questions can be posed by examining the trade-off between the $CV$- criteria described by Eqs. (\ref{cv}) and (\ref{main}). For instance, we can ask: Can resetting favor the conditional exit over the unconditional exits or vice versa? A corollary to this would be when it can be beneficial or detrimental to both. 

To address these questions, we turn our attention to Fig.~\ref{fig1}(c) which showcases a universal phase-diagram spanned by $CV$ and $CV^\sigma$ (the parameters and details are specific to the system studied but not crucial for a generic illustration). While $CV=1$ separates the regimes for the unconditional exits, it can not provide insights for the conditional exits. This is done by drawing the $CV^\sigma=\Lambda^\sigma$ line. These two curves can intersect at a line (or a single point depending on the system) spanned by the coordinates $(1,\Lambda^\sigma_c)$ where 
\begin{equation}\label{lambda-c}
    \Lambda^\sigma_c=\dfrac{\langle T_0\left(\mathbf{\Sigma}\right)\rangle}{\langle T^{\sigma}_0\left(\mathbf{\Sigma}\right)\rangle},
\end{equation} 
which is illustrated with white dashed line in the Venn diagram in Fig.~\ref{fig1}(c). 
The Venn diagram in \fref{fig1}(c) underpins various trade-offs by identifying four different regimes, namely
\begin{itemize}
    \item Region I where $CV^\sigma>\Lambda^\sigma$ and $CV>1$ so that resetting is beneficial to both conditional and unconditional exits
    \item Region II where $CV^\sigma>\Lambda^\sigma$ and $CV<1$ so that resetting is beneficial to  conditional, but detrimental to unconditional exits;
    \item Region III where $CV^\sigma<\Lambda^\sigma$ and $CV<1$ so that resetting is detrimental to both conditional and unconditional exits
    \item Region IV where $CV^\sigma<\Lambda^\sigma$ and $CV>1$ so that resetting is detrimental to conditional exits, but beneficial to unconditional
\end{itemize}
Thus, exploring the phase diagram with the variation of system specific parameters allows us to choose preferential outcomes by imposing resetting.

\subsection{Preferential biasing mediated by resetting}
\label{preferential}
Quite interestingly, the $CV^\sigma$ criterion can also serve as a control to bias the system among a set of conditional escapes. To illustrate this, let us consider two specific outcomes: $\sigma_1$ (desired) and $\sigma_2$ (undesired) out of all the possibilities. For given initial configurations and system parameters (captured in $\mathbf{\Sigma}$), \eref{main} essentially states that to favor the desired outcome, one should have 
\begin{align}
    CV^{\sigma_1}>\Lambda^{\sigma_1}~~\text{and}~~~CV^{\sigma_2}<\Lambda^{\sigma_2},
    \label{eq:preferential}
\end{align}
simultaneously as a sufficient condition. This argument can also be generalized to mutually exclusive sets namely $\{ \sigma_1 \}$ (desired set) and $\{ \sigma_2 \}$ (undesired set). Evidently, the condition for resetting to favor two desired sets should be $CV^{\{\sigma_1 \}}>\Lambda^{\{ \sigma_1 \} }, CV^{\{\sigma_2 \}}>\Lambda^{\{ \sigma_2 \} }$.

To illustrate the universal conditions delineated in this section, we study a diffusive process in a 1D box with two exit points. This is done in the next section.



\begin{figure*}[t!]
    \centering
    \includegraphics[width=0.7\linewidth]{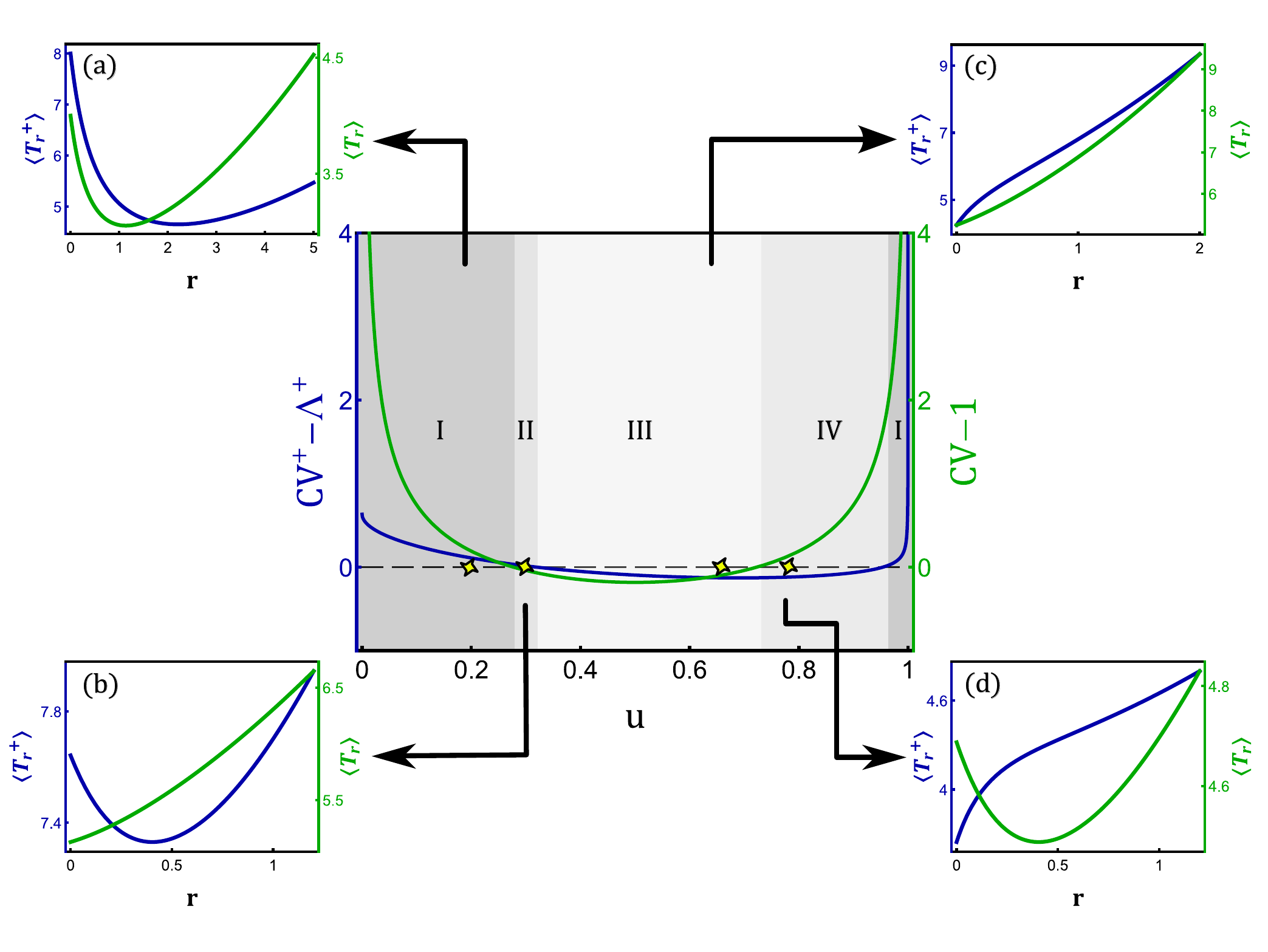}
    \caption{Phase diagram demonstrating the various regions of resetting based optimization for the conditional and unconditional mean exit times in the diffusion problem of Sec. \ref{sec:diffusion}. The control parameter is $u$ and we compare $\langle T_r^+\rangle$ (conditional exit through the right) and $\langle T_r\rangle$ (unconditional exit averaged over both boundaries).  The solid blue curve indicates $CV^+-\Lambda^+$ and the green one $CV-1$ as a function of $u$. Depending on the intersecting points with the zero line (dashed horizontal line), four distinct regions are defined as discussed in Sec.~\ref{formalism:trade-off}. Region I spans the domain of $u$ for which $CV^+>\Lambda^+~\&~CV>1$ and thus resetting optimizes both conditional and unconditional mean exit times, as can also be seen from panel (a) where we have plotted $\langle T_r^+\rangle$ and $\langle T_r\rangle$ as a function of $r$ (for a value of $u$ indicated by a plot-marker in this region). In Region II, $u$ is such that $CV^+>\Lambda^+$ but $CV<1$, hence a finite resetting rate optimizes $\langle T_r^+\rangle$ but not $\langle T_r\rangle$, see panel (b). Region III represents the case where both $CV^+<\Lambda^+$ and $CV<1$, hence resetting is detrimental for both exit times, see panel (c). Finally, in Region IV, $CV^+<\Lambda^+$ but  $CV>1$, therefore the unconditional process is benefited with the introduction of resetting, but not the conditional exit through the right boundary. This is confirmed from panel (d). 
    }
    \label{fig2}
\end{figure*}

\section{Diffusion under resetting}
\label{sec:diffusion}
Let us consider a Brownian particle diffusing inside a one dimensional box of length $l$. The particle starts off at $x_0$ and is reset to the same location at a rate $r$. The box has two exit points, at $0$ and  $l$, from which the particle can escape the interval; thus, in this case $\{\sigma\}$ has two elements, which we denote as $\{ +,- \}$, where + (-) denotes the right (left) boundary. This problem was extensively studied in \cite{pal2019first} where exact expressions for the conditional and unconditional exit times were derived (see also  Appendix \ref{appendixC}). Following ~\cite{pal2019first}, the unconditional mean escape time $\langle T_r \rangle$ of the particle from the interval reads
\begin{equation}\label{mfpt-res}
    \langle T_r(u)\rangle=\dfrac{1}{r}\left[\dfrac{\cosh \left(\dfrac{\alpha_0 l}{2}\right)}{\cosh\left(\dfrac{1}{2}\alpha_0 l(1-2u)\right)}-1\right],
\end{equation}
where $u=x_0/l ~(0 \leq u \leq 1)$ is the dimensionless starting position and $\alpha_0=\sqrt{r/D}$, bearing the dimension of inverse length, is the typical distance covered by the particle between two resetting events. On the other hand, the conditional mean escape times through the right/left exit points, can be written 
from Eq.~\eqref{conditional-mfpt} in the following way
\begin{eqnarray}
    \langle T_r^+\rangle = \langle T_r\rangle - \dfrac{1}{r}\dfrac{1}{e^{2ul\alpha_0}-1}F(u,l),\label{cond-fp-res}\\
    \langle T_r^-\rangle = \langle T_r\rangle + \dfrac{1}{r}\dfrac{e^{l\alpha_0}}{e^{2l\alpha_0}-e^{2ul\alpha_0}}F(u,l),
    \label{cond-fp-resb}
\end{eqnarray}
where
\begin{equation}
    F(u,l)=\dfrac{l\alpha_0\left[e^{2l\alpha_0}-e^{4ul\alpha_0}+(1-2u)e^{2ul\alpha_0}(1-e^{2l\alpha_0})\right]}{2(e^{l\alpha_0}-1)(e^{l\alpha_0}+e^{2ul\alpha_0})}\nonumber.
\end{equation}
For resetting to expedite the unconditional exit time, it is sufficient to utilize the criterion \eqref{cv} where $CV(u)=\sqrt{\dfrac{1-2u+2u^2}{3(u-u^2)}}$ for the underlying process. 
This determines the domain in which resetting expedites the completion of the underlying process: $\mathcal{D}=[ (0, u_{-}) ~\cup~ (u_{+},1) ]$, where $u_{\pm}=(5\pm \sqrt{5})/10$. Hence, if the particle starts closer to either of the boundaries within the threshold $u_\pm$, the completion will be accelerated \cite{pal2019first}. 

The criteria for the conditional exits are more convoluted and require the $CV^{\pm}$ for the underlying process. Following  Appendix \ref{appendixC}, we can write these quantities along with their corresponding thresholds

{\footnotesize
\begin{align}
    &CV^+\left(u\right)=\sqrt{
    \dfrac{2(1+u^2)}{5(1-u^2)}
    },~\Lambda^+\left(u\right)=\sqrt{\dfrac{3u(1+u-u^2)}{2(1-u)(1+u)^2}},
    \label{cv-plus}\\
    &CV^-\left(u\right)=\sqrt{\dfrac{2(u^2-2u+2)}{5u(2-u)}},~\Lambda^-\left(u\right)=\sqrt{\dfrac{3(1-2u^2+u^3)}{2u(2-u)^2}}.\label{cv-minus}
\end{align}}
In what follows, we will use these expressions to discuss various trade-offs highlighted in the preceding section.

\subsection{Resetting favors the conditional escape}
For resetting to expedite the conditional exit times compared to the same of the underlying process, one should satisfy the criterion  \eqref{main} and identify the regions of $u$ correspondingly. 
To see this, 
we first set $CV^\pm\left(u\right)=\Lambda^\pm\left(u\right)$ 
which provide us the limiting points for the right and left exit boundaries respectively
\begin{eqnarray}
    4-11 u-11u^2+19u^3=0\label{cv-plus-cond},\\
    1-24u+46u^2-19u^3=0\label{cv-minus-cond}.
\end{eqnarray}
Solving those two equations, we identify the respective domains namely: (i) \textbf{$CV^+>\Lambda^+$} if $u\in\left(0,0.318051\right)\cup\left(0.954428,1\right)$, and symmetrically (ii)
\textbf{$CV^->\Lambda^-$} if $u\in\left(0,0.0455723\right)\cup\left(0.681949,1\right)$. The former condition is illustrated in Fig. \ref{fig2} which displays the different regions for $\sigma=+$ and typical shapes of the mean exit times  $\langle T_r^+ \rangle$ and $\langle T_r \rangle$ vs. $r$ depending on the starting position $u$. The  regions have the same nomenclature as in Fig. \ref{fig1}. Resetting favors conditional escape to the right boundary in regions I and II and there exists an optimal resetting rate for that quantity. On the contrary, it is observed that resetting prolongs the conditional exit to the right for the central part of the domain \textit{i.e.}, regions III and IV, where $\langle T^+_r\rangle$ is monotonously increasing with $r$. A similar analysis can also be made for $\sigma=-$, the exit time through the left boundary. For brevity, we have relegated this discussion to Appendix \ref{sec:conditional_left}. 

The decrease of a mean \textcolor{black}{first-passage} time with a small $r$ is commonly explained by the fact that resetting eliminates long trajectories wandering away from an initially nearby target state. In an interval, resetting thus reduces the unconditional exit time $\langle T_r\rangle$ when the starting position is sufficiently close to either absorbing boundary, i.e., $u$ close to 0 or 1 \cite{pal2019first}. The small $u$ regions I and II of Fig. \ref{fig2} may therefore look surprising, as resetting, by bringing the particle back close to the left boundary, reduces the conditional time needed to reach the opposite boundary to the right. This paradox can be resolved by noticing that in the absence of resetting, the trajectories contributing to $\langle T^+_0\rangle$ in the limit $u\ll 1$ can be of two different types: some trajectories $x(t)$ wander for a long time, creating loops, in the central part of the domain before reaching $u=1$ 
(let us denote them as trajectories of type $L_0$), while other trajectories (say, of type $D_0$) are much more direct in crossing the interval. At small $r$, a $L_0$-trajectory is more likely to undergo a reset, while a $D_0$ trajectory manages to reach the right boundary before the first resetting event. A $L_0$-trajectory which has been reset is then likely to be absorbed at the origin, as $\epsilon_0^{-}(u\ll1)=1-u\sim 1$ [see \eref{eq:exit_prob}], and will no longer contribute to $\langle T^+_r\rangle$. Hence the long trajectories that initially contributed to $\langle T^+_0\rangle(u\ll 1)$ tend to be eliminated by resetting and the rare but fast $D_0$-trajectories are much less affected, resulting in a overall decrease of $\langle T^+_r\rangle(u\ll 1)$.

In the central part of the domain ($u\sim 1/2$) the situation is different because $\epsilon_0^{-}(u)\sim 1/2$: a $L_0$-trajectory that is reset once still has a large probability to reach the right boundary and to contribute to $\langle T^+_r\rangle$. In this case, resetting prolongs the time spent by the particle in the interval.

\subsection{The tug-of-war between conditional and unconditional exits}\label{cond-uncond-tradeoff}
We now turn our attention to elucidate the trade-offs between conditional and unconditional times as outlined in Sec. \ref{formalism:trade-off}. The idea is to identify the domains of $u$ for which one can see positive effects of resetting on the conditional times over the unconditional ones, and vice-versa. 

Region I is defined by $CV^+>\Lambda^+~\cap~CV>1$, and thus resetting is beneficial for both the mean unconditional time and mean conditional escape time from the right boundary. This inequality is valid for  the initial condition $u\in \biggl(0,\frac{5-\sqrt{5}}{10}\biggl)\cup \biggl(0.954428,1\biggl)$, 
as can also be seen in Fig.~\ref{fig2}(a). Region II represents \textbf{$CV^+>\Lambda^+~\cap~CV<1$} which results in $u\in \biggl(\frac{5 - \sqrt{5}}{10},0.318051\biggl)$. In this domain, the unconditional mean time cannot be optimized by resetting but the conditional time to the right boundary can be optimized, as depicted in Fig~\ref{fig2}(b). To define the region III, we set $CV^+<\Lambda^+~\cap~CV<1$ so that resetting becomes detrimental in optimizing both conditional and unconditional escape time. This is evident from Fig.~\ref{fig2}(c) for $u \in \biggl(0.318051,\frac{5+\sqrt{5}}{10}\biggl)$. Finally, one finds region IV by imposing \textbf{$CV^+<\Lambda^+~\cap~CV>1$} resulting in $u\in \biggl(\frac{5 + \sqrt{5}}{10},0.954428\biggl)$. In this case,  resetting turns out to be helpful for unconditional exit albeit prolonging the conditional exit from the right, as depicted in Fig.~\ref{fig2}(d). Similarly, we can compare the optimization of conditional escape time from the left boundary with the unconditional escape time -- this is demonstrated in Appendix~\ref{sec:conditional_left}.

\subsection{Preferential biasing between the right and left exit}
\begin{figure}[t!]
    \centering
    \includegraphics[width=\linewidth]{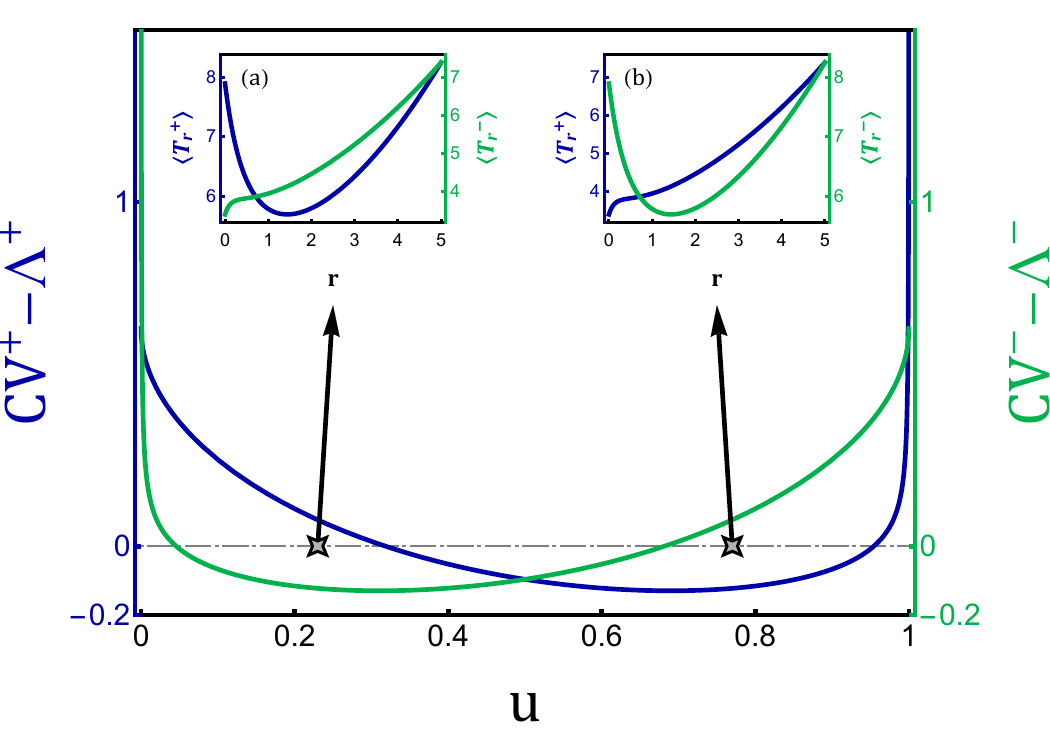}
    \caption{Illustration of the preferential biasing between the left and right exit for the diffusion process by leveraging the criterion [see \eref{eq:preferential} and \eref{eq:preferential_diffusion}]. We plot $CV^+-\Lambda^+$ (blue solid line) and $CV^- -\Lambda^-$ (green solid line) as a function of $u$. For the region of $u$ where $CV^+>\Lambda^+$ but $CV^-<\Lambda^-$, resetting expedites the right exit but not the left, as can be seen from inset (a) showing $\langle T_r^+\rangle$ vs. $r$ (which can be optimized with $r$) and  $\langle T_r^-\rangle$ (which cannot). In the symmetric region of $u$, we have $CV^+<\Lambda^+$ but $CV^->\Lambda^-$ so that resetting becomes beneficial for the left exit but not for the right exit. This is also confirmed by the variations of the mean conditional times with $r$ in inset (b). 
  }
  \label{fig5}
\end{figure}

In Sec. (\ref{formalism:trade-off}), we discussed how one can probe resetting to bias between multiple outcomes. In the case of diffusion, this implies that one can prefer the bias for the right exit prior to the left exit or vice-versa. For instance, to bias the right exit (i.e., to expedite completion for processes that exit through the right) by applying resetting, we should identify the regimes for $u$ such that 
\begin{align}
    CV^+(u)>\Lambda^+(u)~~\text{and}~~CV^-(u)<\Lambda^-(u).
    \label{eq:preferential_diffusion}
\end{align}
This results in the following domain of initial location $u\in \left(0.0455723,0.318051\right)$. A similar analysis holds if we want to bias the conditional time through the left boundary in comparison to the right. For such case,  one can identify the following domain
$u\in \left(0.681949,0.954428\right)$. 
Fig.~\ref{fig5} showcases all these scenarios together. 

\section{Conclusions}
We have examined a generic \textcolor{black}{first-passage} process performing a task under resetting dynamics and furthermore, possessing multiple exit points or outcomes. The task is said to be completed when the process finishes through any of these exit pathways. Over many realizations, the exit pathways can vary as the process is stochastic. The unconditional \textcolor{black}{first-passage} time accounts for exit time averaged over all such pathways while the conditional \textcolor{black}{first-passage} time is sampled over only those pathways that complete via a particular exit point of interest. Over the last decade, a persistent effort has been made to understand the effects of resetting on both observables and various optimization properties have been studied across many different systems. 

One notable result in the field is the so-called $CV$-criterion that serves as a sufficient condition for resetting to the initial condition to improve the speed of a \textcolor{black}{first-passage} process \cite{pal2024random}. However, this is pertinent only to the mean unconditional time, and not to the conditioned one. Bridging this gap, we have discovered a universal $CV^\sigma$-criterion that governs the efficiency of a conditional outcome. The condition is seen to be sufficient  to mark the speed-up over a resetting free process conditioned on one or a set of preferred outcomes. Leveraging this further, we have shown how resetting strategy can be used to bias between various conditional outcomes, in particular between the selective and non-selective ones. Quite interestingly, the validity of the criterion is highly sensitive to the magnitude of the fluctuations for the conditional time emanating from the resetting-free process. \textcolor{black}{Finally, the criterion is universal as it does not depend on the nature of the stochastic process, dimensionality or target distribution as long as targets are finite in numbers and well-separated to be identified as distinct regions}. Nonetheless, when applying the 
 $CV^\sigma$-criterion, it is important to note that it assumes the underlying search process—without resetting—has finite mean and variance for the first-passage time. In unconfined search scenarios, this condition may not hold, rendering the criterion pathological, as it would then predict that resetting is always beneficial, regardless of context.

To demonstrate \textcolor{black}{how to use the relation in practice,} 
we have examined it on a one dimensional diffusion process with two exit points. As a non-intuitive result, we have found that it is possible to optimize a conditional exit time via resetting by choosing the starting position {\it far} from that exit and close to another exit. This feature is opposite to what observed with unconditional times, where resetting typically helps to find close-by targets. This phenomenon is due to the fact that long excursions ending at the further exit in the absence of resetting, thus contributing to large conditional times, get eliminated when resetting is switched on. Meanwhile, more direct trajectories to the further exit are preserved. Interestingly, resetting is  also seen to improve a conditional exit time if the diffusive process starts close to that exit, but at a distance much shorter than expected for the unconditional problem.



Processes with multiple outcomes are abundant in nature starting from \textcolor{black}{the localization of a specific site by a protein on a DNA \cite{coppey2004kinetics},} gated chemical reactions \cite{biswas2023rate,scher2024continuous}, enzymatic reactions \cite{reuveni2014role}, channel facilitated transport \cite{jain2023fick,pal2024channel}, directed intermittent search in cellular biology such as cytoneme based morphogenesis \cite{bressloff2020directed}, motor driven intracellular transport  \cite{bressloff2020modeling} and in artificial systems such as queues, algorithms and games. Many such systems have resetting integrated to their dynamics either intrinsically or externally and thus, we believe that our results hold severe importance in harnessing the nature of the outcome. Furthermore, the universality of these results should be applicable to a very broad category of stochastic search systems. An immediate avenue to test our results would be a directed search process in higher dimensions in the presence of multiple targets or exit points \cite{evans2014diffusion,holcman2014narrow,metzler2014first,bray2013persistence}. 
Finally, our universal results could also be verified using optical trap experiments \cite{tal2020experimental,besga2020optimal,goerlich2023experimental} or light controlled robots \cite{paramanick2024uncovering}.\\


\section{Acknowledgment}
The numerical calculations reported in
this work were carried out on the Nandadevi and Kamet cluster, which are maintained and supported by the Institute of Mathematical Science’s High-Performance Computing Center.  AP gratefully acknowledges research support from the Department of Atomic Energy, Government of India via the Apex Projects. AP acknowledges the support from the International Research Project (IRP) titled ``Classical and quantum dynamics in out of equilibrium systems'' by CNRS, France. The authors thank the Higgs Center for Theoretical Physics, Edinburgh, for hospitality during the workshop ``New Vistas in Stochastic Resetting'' where several discussions related to the project took place.







\appendix

\section{Derivation of \eref{conditional-mfpt}}\label{appendixA}
In this section, we provide the derivation of Eq.~\eqref{epsilon-2}. We start by recalling
Eq.~\eqref{def-conditional-mfpt} which is the mean conditional \textcolor{black}{first-passage}
\begin{equation}\label{A1}
    \langle T^{\sigma}_r(\mathbf{\Sigma})\rangle=\dfrac{\int_{0}^{\infty} dt~t~J^{\sigma}_r(\mathbf{\Sigma},t)}{\int_{0}^{\infty} dt~J^{\sigma}_r(\mathbf{\Sigma},t)},
\end{equation}
written in terms of the currents. 
Taking the Laplace transform, we can rewrite \eref{A1} so that
\begin{equation}\label{A2}
    \langle T^{\sigma}_r(\mathbf{\Sigma})\rangle=\dfrac{\left.-\dfrac{\partial\widetilde{J}_r^{\sigma}(\mathbf{\Sigma},s)}{\partial s}\right|_{s\rightarrow 0}}{\widetilde{J}_r^{\sigma}(\mathbf{\Sigma},s\rightarrow 0)}=-\left.\dfrac{\partial}{\partial s}\ln{\widetilde{J}^{\sigma}_r(\mathbf{\Sigma},s)}\right|_{s\rightarrow 0}.
\end{equation}
Using the renewal relation in Eq.~\eqref{renewal-lap-J} for the current, we can rewrite the above equation further
\begin{align}\label{A3}
    \langle T^{\sigma}_r(\mathbf{\Sigma})\rangle=&-\left.\dfrac{\partial}{\partial s}\ln{\left[\widetilde{J}^{\sigma}_0(\mathbf{\Sigma},r+s)\right]}\right|_{s\rightarrow 0}\nonumber\\
    &+\left.\dfrac{\partial}{\partial s}\ln{\left[1-r\widetilde{Q}_0(\mathbf{\Sigma},r+s)\right]}\right|_{s\rightarrow 0}.
\end{align}
We now use the relation $\widetilde{J}^\sigma_0\left(\mathbf{\Sigma},s\right)=\epsilon^{\sigma}_0\left(\mathbf{\Sigma}\right)\widetilde{T}^\sigma_0\left(\mathbf{\Sigma},s\right)$ from \eref{FPT_r} and \eref{epsilon-2} in \eref{A3} to further simplify. 
In addition, we also make use of the following identity
\begin{align}
    \frac{\partial}{\partial s}  \widetilde{T}_0^\sigma(r+s)=\frac{\partial}{\partial s}  \left\langle e^{-(r+s)T_0^\sigma}\right \rangle=\frac{\partial}{\partial r}  \left\langle e^{-(r+s)T_0^\sigma}\right \rangle    
\end{align}
so that $ \frac{\partial}{\partial s}  \widetilde{T}_0^\sigma(r+s)\bigg|_{s \to 0}=\frac{\partial}{\partial r}  \widetilde{T}_0^\sigma(r)$ and thus,
 the first term on the RHS of Eq.~\eqref{A3} can be rewritten in the following way
\begin{align}
    -\left.\dfrac{\partial}{\partial s}\ln{\left[\widetilde{J}^{\sigma}_0(\mathbf{\Sigma},r+s)\right]}\right|_{s\rightarrow 0}&=-\dfrac{\left.\dfrac{\partial}{\partial s}\left[\widetilde{J}^\sigma_0\left(\mathbf{\Sigma},r+s\right)\right]\right|_{s\rightarrow 0}}{\left[\widetilde{J}^\sigma_0\left(\mathbf{\Sigma},r\right)\right]}\nonumber\\
    &=-\dfrac{\dfrac{\partial}{\partial r}\widetilde{T}^\sigma_0\left(\mathbf{\Sigma},r\right)}{\widetilde{T}^\sigma_0\left(\mathbf{\Sigma},r\right)}\nonumber\\
    &=-\dfrac{\partial}{\partial r}\ln\left[\widetilde{T}^\sigma_0\left(\mathbf{\Sigma},r\right)\right].
    \label{eq:A5}
\end{align}

The second term on the right-hand side of Eq.~\eqref{A3} can be simplified in a similar manner yielding to
\begin{align}\label{A5}
    &\left[\dfrac{\partial}{\partial s}\ln{\left[1-r\widetilde{Q}_0(\mathbf{\Sigma},r+s)\right]}\right]_{s\rightarrow 0}=\nonumber\\
    &\dfrac{r}{1-r\widetilde{Q}_0(\mathbf{\Sigma},r)}\left[-\dfrac{\partial }{\partial r}\widetilde{Q}_0(\mathbf{\Sigma},r)\right].
\end{align}
Using the relation between the \textcolor{black}{first-passage} time density and the survival probability in Laplace frame \textit{i.e.}, $\widetilde{T}_0\left(r,\mathbf{\Sigma}\right)=1-r\widetilde{Q}_0\left(\mathbf{\Sigma},r\right)$, we can rewrite the above equation in the following way
\begin{align}\label{A7}
    &\left[\dfrac{\partial}{\partial s}\ln{\left[1-r\widetilde{Q}_0(\mathbf{\Sigma},r+s)\right]}\right]_{s\rightarrow 0}=\nonumber\\
    &\dfrac{1-\widetilde{T}_0\left(\mathbf{\Sigma},r\right)}{r\widetilde{T}_0\left(\mathbf{\Sigma},r\right)}+\dfrac{\partial}{\partial r}\ln\left[\widetilde{T}_0\left(\mathbf{\Sigma},r\right)\right].
\end{align}
Substituting \eref{eq:A5}, \eref{A7} and the unconditional mean \textcolor{black}{first-passage} time in the presence of resetting given by $\langle T_r\left(\mathbf{\Sigma}\right)\rangle=\widetilde{Q}_0\left(\mathbf{\Sigma},r\right)/\widetilde{T}_0\left(\mathbf{\Sigma},r\right)$ into Eq.~\eqref{A3} we have
\begin{equation}
    \langle T^{\sigma}_r\left(\mathbf{\Sigma}\right)\rangle = \langle T_r\left(\mathbf{\Sigma}\right)\rangle+\dfrac{\partial}{\partial r}\ln\left[\dfrac{\widetilde{T}_0\left(\mathbf{\Sigma},r\right)}{\widetilde{T}^\sigma_0\left(\mathbf{\Sigma},r\right)}\right]\nonumber.
\end{equation}
which is Eq.~\eqref{conditional-mfpt} in the main text.

\section{Derivation of \eref{conditional-cv}}\label{appendixB}
In this section, we derive the relation, depicted in Eq.~\eqref{conditional-cv}. To this end, we rewrite \eref{conditional-mfpt} in the following way
\begin{equation}\label{A10}
    \langle T^{\sigma}_r\left(\mathbf{\Sigma}\right)\rangle = \langle T_r\left(\mathbf{\Sigma}\right)\rangle+\left(\dfrac{\partial_r\widetilde{T}_0\left(\mathbf{\Sigma},r\right)}{\widetilde{T}_0\left(\mathbf{\Sigma},r\right)}-\dfrac{\partial_r\widetilde{T}^{\sigma}_0\left(\mathbf{\Sigma},r\right)}{\widetilde{T}^{\sigma}_0\left(\mathbf{\Sigma},r\right)}\right).
\end{equation}
To derive the criterion, let us consider an underlying \textcolor{black}{first-passage} process, introduce an infinitesimal resetting rate $\delta r \to 0$ and see its effect. To proceed further, we do a Taylor series expansion of $\widetilde{T}_0(r)$ and $\widetilde{T}^{\sigma}_0(r)$ in the vicinity of $\delta r \to 0$, so that 
\begin{align}\label{eq:T_0-Taylor}
    \widetilde{T}_0(\delta r)&=1-\delta r\langle T_0\rangle+\frac{\delta r^2}{2}\langle T_0^2\rangle-\cdots, \nonumber \\
    \widetilde{T}^{\sigma}_0(\delta r)&=1-\delta r\langle T^{\sigma}_0\rangle+\frac{\delta r^2}{2}\langle (T^{\sigma}_0)^2\rangle-\cdots
\end{align}
Substituting Eqs. from (\ref{eq:T_0-Taylor}) into \eref{A10}
\begin{align}
    \langle T^{\sigma}_{\delta r}\rangle=&\langle T_{\delta r}\rangle+\bigg(\dfrac{-\langle T_0\rangle+\delta r\langle T_0^2\rangle-\cdots}{1-\delta r\langle T_0\rangle+\frac{\delta r^2}{2}\langle T_0^2\rangle-\cdots}\nonumber\\
    &-\dfrac{-\langle T^{\sigma}_0\rangle+\delta r\langle (T^{\sigma}_0)^2\rangle-\cdots}{1-\delta r\langle T^{\sigma}_0\rangle+\frac{\delta r^2}{2}\langle (T^{\sigma}_0)^2\rangle-\cdots}\bigg).
    \label{eq:expansion_2}
\end{align}
Further noting $\langle T_{\delta r}\rangle=\langle T_{0}\rangle+\delta r\left(\langle T_{0}\rangle^2-\dfrac{1}{2}\langle T_{0}^2\rangle\right)+\cdots$, we arrive at the following representation of \eref{eq:expansion_2} 
\begin{equation}\label{B1}
    \langle T^{\sigma}_{\delta r}\rangle=\langle T^{\sigma}_0\rangle+a_1~\delta r+a_2~\delta r^2+O(\delta r^3),
\end{equation}
where we identify $a_1=\dfrac{\langle T_0^2\rangle}{2}-\langle (T^{\sigma}_0)^2\rangle+\langle T^{\sigma}_0\rangle^2$ and $a_2=\dfrac{1}{6}\left(6\langle T^{\sigma}\rangle^3-9\langle T^{\sigma}\rangle \langle (T^{\sigma})^2\rangle+3\langle (T^{\sigma})^3\rangle+3\langle T_0\rangle\langle T_0^2\rangle-2\langle T_0^3\rangle\right)$. We can further simplify $a_1$ in the following way
\begin{align}
    a_1&=\dfrac{\langle T_0^2\rangle}{2}-\langle (T^{\sigma}_0)^2\rangle+\langle T^{\sigma}_0\rangle^2,\nonumber\\
    &=\dfrac{\langle T_0\rangle^2}{2}(1+CV^2)-\langle T^{\sigma}_0\rangle^2\dfrac{\langle (T^{\sigma}_0)^2\rangle-\langle T^{\sigma}_0\rangle^2}{\langle T^{\sigma}_0\rangle^2},\nonumber\\
    &=\dfrac{\langle T_0\rangle^2}{2}(1+CV^2)-\langle T^{\sigma}_0\rangle^2 \left(CV^{\sigma}\right)^2,\nonumber\\
    &=\langle T^{\sigma}_0\rangle^2 \left[\dfrac{\langle T_0\rangle^2}{2\langle T^{\sigma}_0\rangle^2}(1+CV^2)-\left(CV^{\sigma}\right)^2\right],\nonumber\\
    &=\langle T^{\sigma}_0\rangle^2\left[(\Lambda^{\sigma})^2-(CV^{\sigma})^2\right],\nonumber
\end{align}
where we have defined
\begin{align}
    CV^{\sigma}&=\sqrt{\dfrac{\langle (T^{\sigma}_0)^2\rangle-\langle T^{\sigma}_0\rangle^2}{\langle T^{\sigma}_0\rangle^2}},\\
    \Lambda^{\sigma}&=\sqrt{\dfrac{\langle T_0\rangle^2}{2\langle T^{\sigma}_0\rangle^2}(1+CV^2)}.
\end{align}
Keeping upto the linear response term, Eq.~\eqref{B1} can be rewritten as
\begin{equation}
    \langle T^{\sigma}_{\delta r}\rangle=\langle T^{\sigma}_0\rangle+\delta r \langle T^{\sigma}_0\rangle^2\left[(\Lambda^{\sigma})^2-(CV^{\sigma})^2\right].
    \label{eq:criterion}
\end{equation}
For resetting to expedite the mean conditional time, one should have $\langle T^{\sigma}_{\delta r}\rangle<\langle T^{\sigma}_{0}\rangle$ 
which results in, from \eref{eq:criterion}, the criterion $CV^{\sigma}>\Lambda^{\sigma}$ that is the central result of our letter, announced in the main text as \eref{conditional-cv}.

\section{Moments of conditional first-passages for the resetting free underlying diffusion problem}\label{appendixC}
The 
Fokker-Planck equation for the probability density function of a diffusing particle is given by \cite{redner2001guide}
\begin{equation}\label{C1}
    \dfrac{\partial }{\partial t}p_0(x,t)=D\dfrac{\partial^2}{\partial x^2}p_0(x,t).
\end{equation}
To implement the absorbing boundary conditions at $x=0$ and $x=l$, one should have $p_0(0,t)=0,~p_0(l,t)=0$. The initial condition is fixed and thus $p_0(x,0)=\delta(x-x_0)$. To solve for the currents, we need to solve Eq.~\eqref{C1} which can be done in the Laplace space following
\begin{equation}
    s~\widetilde{p}_0(x,s)-\delta(x-x_0)=D\dfrac{\partial^2}{\partial x^2}\widetilde{p}_0(x,s),\nonumber
\end{equation}
such that the solution for the propagator is given by
\begin{align}
    \widetilde{p}_0(x,s)=&\Theta(x_0-x)\dfrac{\sinh\left[x\sqrt{\dfrac{s}{D}}\right]\sinh\left[(l-x_0)\sqrt{\dfrac{s}{D}}\right]}{\sqrt{sD}\sinh\left[l\sqrt{\dfrac{s}{D}}\right]}\nonumber\\
    &+\Theta(x-x_0)\dfrac{\sinh\left[(l-x)\sqrt{\dfrac{s}{D}}\right]\sinh\left[x_0\sqrt{\dfrac{s}{D}}\right]}{\sqrt{sD}\sinh\left[l\sqrt{\dfrac{s}{D}}\right]}\nonumber,
\end{align}
where $\Theta(x)=1$ if $x>0$ otherwise $0$ for $x<0$. The above expression is enough to extract the  currents(in Laplace space) through the right and left escape respectively namely
\begin{align}\label{C2}
    \widetilde{J}_0^+(x_0,s)&=-\left.D\dfrac{\partial}{\partial x}\widetilde{p}_0(x,s)\right|_{x\rightarrow l}\nonumber\\
    &=\text{cosech}\left(l\sqrt{\dfrac{s}{D}} \right)\text{sinh} \left(x_0\sqrt{\dfrac{s}{D}} \right),\nonumber\\
    \widetilde{J}_0^-(x_0,s)&=\left.D\dfrac{\partial}{\partial x}\widetilde{p}_0(x,s)\right|_{x\rightarrow 0}\nonumber\\
    &=\text{cosech}\left(l\sqrt{\dfrac{s}{D}} \right)\text{sinh} \left((l-x_0)\sqrt{\dfrac{s}{D}}\right).
\end{align}
The exit probabilities are given by
\begin{align}
\label{eq:exit_prob}
   \epsilon_0^+(u)=\widetilde{J}_0^+(u,0)=u,~~~ \epsilon_0^-(u)=\widetilde{J}_0^-(u,0)=1-u,
\end{align}
where recall that $u=x_0/l$ is the scaled variable. Furthermore, the conditional \textcolor{black}{first-passage} time densities in Laplace space are given by
\begin{align}\label{C3}
    \widetilde{T}_0^+(u,s)&=\dfrac{1}{u}\text{cosech}\left(l\sqrt{\dfrac{s}{D}} \right)\text{sinh} \left(u l\sqrt{\dfrac{s}{D}} \right),\nonumber\\
    \widetilde{T}_0^-(u,s)&=\dfrac{1}{1-u}\text{cosech}\left(l\sqrt{\dfrac{s}{D}} \right)\text{sinh} \left((1- u)l\sqrt{\dfrac{s}{D}}\right).
\end{align}
Finally, the unconditional \textcolor{black}{first-passage} time density can be found by averaging over both the possibilities 
\begin{align}\label{C4}
    \widetilde{T}_0(u,s) &= \epsilon^+_0(u)\widetilde{T}^+_0(u,s)+\epsilon^-_0(u)\widetilde{T}^-_0(u,s)\nonumber\\
    &=\text{cosech}\left(l\sqrt{\dfrac{s}{D}} \right)\bigg[\text{sinh} \left( ul\sqrt{\dfrac{s}{D}}\right)\nonumber\\
    &~~~~~~~~~+\text{sinh} \left((1- u)l\sqrt{\dfrac{s}{D}}\right)\bigg].
\end{align}
Given the distribution, the moments are straightforward to compute via the relation
\begin{eqnarray}
    \langle T_0^m\rangle=(-1)^m\left.\dfrac{d^m\widetilde{T}_0}{ds^m}\right|_{s\rightarrow 0}\nonumber.
\end{eqnarray}
For instance, the first and second moments of the conditional times read
\begin{align}\label{C5}
    &\langle T_0^+\rangle=\dfrac{l^2}{6D}(1-u^2),\nonumber\\
    &\langle \left(T_0^+\right)^2\rangle=\dfrac{l^4}{180D^2}\left(7-10u^2+3u^4\right),\nonumber\\
    &\langle T_0^-\rangle=\dfrac{l^2}{6D}u\left(2-u\right),\\
    &\langle \left(T_0^-\right)^2\rangle=\dfrac{l^4}{180D^2}u(2-u)(4+6u-3u^2).\nonumber
\end{align}
Similarly, the first two moments of the unconditional \textcolor{black}{first-passage} time read
\begin{align}\label{C6}
    \langle T_0(u)\rangle&=\dfrac{l^2}{2D}u(1-u),\nonumber\\
    \langle (T_0(u))^2\rangle&=\dfrac{l^4}{12D^2}u\left(1-2u^2+u^3\right),
\end{align}
and using these moments we find the coefficient of variation ($CV$) to be
\begin{equation}\label{C7}
    CV(u)=\sqrt{\dfrac{\langle (T_0(u))^2\rangle-\langle T_0(u)\rangle^2}{\langle T_0(u)\rangle^2}}=\sqrt{\dfrac{1-2u+2u^2}{3(u-u^2)}}.
\end{equation}
Finally, using the exact results in above we find the conditional $CV$-s as
\begin{align}
    CV^+(u)&=\sqrt{\dfrac{\langle (T_0^+(u))^2\rangle-\langle T_0^+(u)\rangle^2}{\langle T_0^+(u)\rangle^2}}= \sqrt{\dfrac{2(1+u^2)}{5(1-u^2)}},\nonumber\\
    CV^-(u)&=\sqrt{\dfrac{\langle (T_0^-(u))^2\rangle-\langle T_0^-(u)\rangle^2}{\langle T_0^-(u)\rangle^2}}= \sqrt{\dfrac{2(u^2-2u+2)}{5u(2-u)}},\nonumber
\end{align}
and the corresponding $\Lambda$-s are given by
\begin{align}
    &\Lambda^+\left(u\right)=\sqrt{\dfrac{\langle T_0\rangle^2}{2\langle T^{+}_0\rangle^2}(1+CV^2)}=\sqrt{\dfrac{3u(1+u-u^2)}{2(1-u)(1+u)^2}},\nonumber\\
    &\Lambda^-\left(u\right)=\sqrt{\dfrac{\langle T_0\rangle^2}{2\langle T^{-}_0\rangle^2}(1+CV^2)}=\sqrt{\dfrac{3(1-2u^2+u^3)}{2u(2-u)^2}}.\nonumber
\end{align}

\section{Comparison of conditional escape from the left region and the unconditional time}
\label{sec:conditional_left}
\begin{figure*}[!ht]
    \centering
    \includegraphics[width=0.7\linewidth]{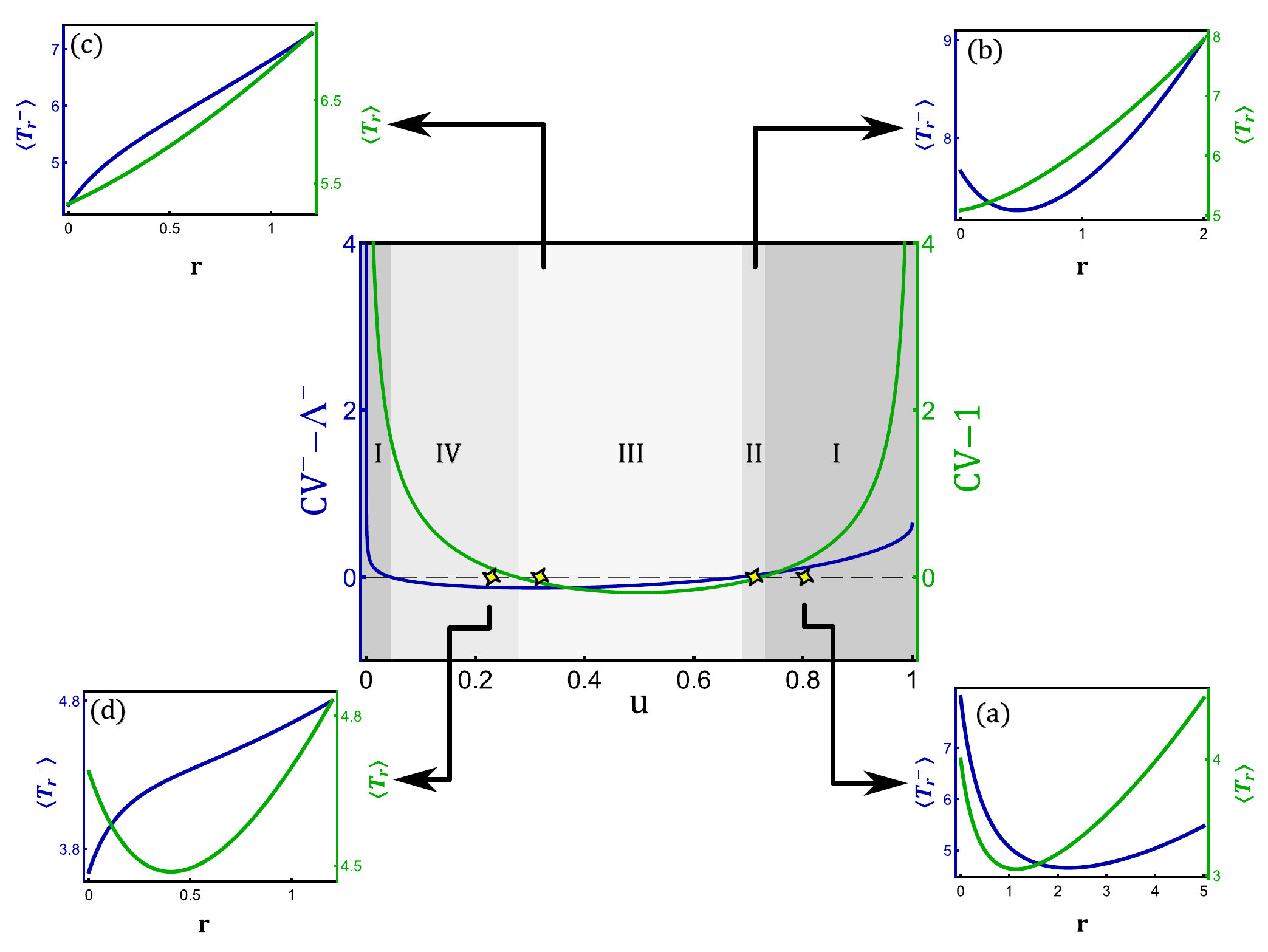}
    \caption{A comparison between conditional escape from left boundary and the unconditional escape. Treating $u$ as a control parameter, we illustrate the behavior of $\langle T_r^-\rangle$ and $\langle T_r\rangle$ in the presence of resetting. The solid blue curve indicates $CV^--\Lambda^-$ and the green one represents $CV-1$ as a function of $u$. Similar to Fig.~\ref{fig2}, we explore four distinct regions of optimization as discussed in Sec.~\ref{formalism:trade-off}. Region I showcases $CV^->\Lambda^-$and $CV>1$ and thus resetting optimizes both conditional and unconditional \textcolor{black}{first-passage} time. It is demonstrated in panel (a), where we have illustrated $\langle T_r^-\rangle$ and $\langle T_r\rangle$ as a function of $r$. Next we explore Region II, where $CV^->\Lambda^-$ but $CV<1$ so that resetting optimizes $\langle T_r^-\rangle$ but not $\langle T_r\rangle$, as shown in panel (b). Region III elaborates a scenario where $CV^-<\Lambda^-$ and $CV<1$ so that resetting prolongs both $\langle T_r^-\rangle$ and $\langle T_r\rangle$ (as shown in panel (c)). Finally, we arrive at the region IV for which $CV^-<\Lambda^-$ but  $CV>1$ so that the unconditional process is benefited with the introduction of resetting, but not the conditional exit from the left boundary. This can also be confirmed from panel (d).}
    \label{fig2-minus}
\end{figure*}

Similar to the discussion made in Sec.~\ref{formalism:trade-off}, here we demonstrate the trade-off between the conditional escape from the left boundary and the unconditional escape in the presence of resetting. For the initial condition $u\in \biggl(0,0.0455723\biggl)\cup \biggl(\frac{5+\sqrt{5}}{10},1\biggl)$ spanned in region I, where $CV^->\Lambda^-~\cap~CV>1$, resetting is beneficial for both conditional escape from the left boundary and unconditional escape as described in Fig.~\ref{fig2-minus}(a). To extract the domain of region II, one should set \textbf{$CV^->\Lambda^-~\cap~CV<1$} which results in $u\in \biggl(0.681949,\frac{5+\sqrt{5}}{10}\biggl)$. In this vicinity, the unconditional mean time cannot be optimized but the conditional time to the left boundary can be optimized, as depicted in Fig~\ref{fig2-minus}(b). 
To find region III, we set $CV^-<\Lambda^-~\cap~CV<1$ revealing $u \in \biggl(\frac{5-\sqrt{5}}{10},0.681949\biggl)$. In here, resetting can not optimize either conditional or unconditional escape, as evident from Fig.~\ref{fig2-minus}(c). Finally, we find region IV by setting \textbf{$CV^-<\Lambda^-~\cap~CV>1$} so that the domain of interest turns out to be $u\in \biggl(0.0455723,\frac{5-\sqrt{5}}{10}\biggl)$. The corresponding plots for the respective MFPTs are displayed in Fig.~\ref{fig2-minus}(d). 

\bibliography{reference}

\end{document}